\documentclass[aip,graphicx, 
reprint,
]{revtex4-1}

\usepackage{graphicx}
\usepackage{dcolumn}
\usepackage{bm}

\usepackage[utf8]{inputenc}
\usepackage[T1]{fontenc}
\usepackage{mathptmx}

\usepackage{xcolor}
\newcommand{\red}[1]{\color{red} \color{black} {#1}}

\newcommand{\sss}{\scriptscriptstyle}
\newcommand{\sst}{\scriptstyle}

\newcommand{\stext}[1]{\sss \rm{#1} \sst}

\usepackage{eso-pic}

\AddToShipoutPicture{
	\raisebox{2\baselineskip}{\makebox[\paperwidth]{\begin{minipage}[t][0.5cm][b]{21cm}\centering \scriptsize The following article has been submitted to the Journal of Vacuum Science \& Technology B. After it is published, it will be found at \texttt{https://avs.scitation.org/journal/jvb}.\end{minipage}
}}}

\begin{document}

\preprint{AIP/123-QED}

\title{THz optical properties of polymethacrylates after thermal annealing}

\author{S.~Park}
 \email{spark71@uncc.edu.}
 \affiliation{Department of Physics and Optical Science, University of North Carolina at Charlotte, 9201 University City Blvd, Charlotte, NC, 28223}
 
\author{Y.~Li}
\affiliation{Department of Physics and Optical Science, University of North Carolina at Charlotte, 9201 University City Blvd, Charlotte, NC, 28223}

\author{D.B.~Fullager}
\affiliation{Laser Tel, 7775 N Casa Grande Highway, Tucson, AZ, 85743}

\author{M.~Lata}
\affiliation{Department of Physics and Optical Science, University of North Carolina at Charlotte, 9201 University City Blvd, Charlotte, NC, 28223}

\author{P.~K{\"u}hne}
\affiliation{THz Materials Analysis Center, Department of Physics, Chemistry, and Biology (IFM), Link{\"o}ping University, SE 581 83 Link{\"o}ping, Sweden}

\author{V.~Darakchieva}
\affiliation{THz Materials Analysis Center, Department of Physics, Chemistry, and Biology (IFM), Link{\"o}ping University, SE 581 83 Link{\"o}ping, Sweden}

\author{T.~Hofmann}
\affiliation{Department of Physics and Optical Science, University of North Carolina at Charlotte, 9201 University City Blvd, Charlotte, NC, 28223}
\affiliation{THz Materials Analysis Center, Department of Physics, Chemistry, and Biology (IFM), Link{\"o}ping University, SE 581 83 Link{\"o}ping, Sweden}
\date{\today}

\begin{abstract}
\noindent Polymer based stereolithographic additive manufacturing has been established for the rapid and low-cost fabrication of THz optical components due to its ability to construct complex 3D geometries with high resolution. For polymer based or integrated optics, thermal annealing processes are often used to optimize material properties. However, despite the growing interest in THz optics fabricated using stereolithography, the effects of thermal annealing on the THz dielectric properties of polymethacrylates compatible with stereolithography has not been studied yet. In this manuscript we report on the THz ellipsometric response of thermally annealed polymethacrylates prepared using UV polymerization. Our findings indicate that the investigated polymethacrylate maintain a stable optical response in THz spectral range from 650 to 950~GHz after thermal annealing at temperatures up to 70~$^{\circ}$C for several hours.
\end{abstract}

\maketitle

\section{\label{sec:level1}Introduction}

Fabrication of THz optical components using additive manufacturing methods has been reported as an efficient alternative to long established fabrication routes using molding or single-point diamond machining, for instance.\cite{squires20153d,Weidenbach:16,Kaur2015} 
\red{Fabrication of THz optical components using additive manufacturing methods has been reported as an efficient alternative to long established fabrication routes using molding or single-point diamond machining, for instance.\cite{squires20153d,Weidenbach:16,Kaur2015} THz optical components have been primarily fabricated using fused filament deposition, which allows the deposition of a wide range of THz transparent materials. The main advantage of fused filament deposition is the lower instrument and material cost. However, fused filament deposition suffers from a low spatial resolution and surface finish, which is mainly caused by the nozzle diameter.\cite{Yan1996} Stereolithography, on the other hand, has been reported to accomplish spatial resolutions on the order of 10~$\mu$m. In addition, the surface finish of components fabricated by stereolithography is substantially better compared to other additive manufacturing fabrication techniques. \cite{NGO2018172,Shallan2014chem}}

\red{Recently, metalization of stereolithographically fabricated parts was demonstrated to be an effective way of prototyping reflective THz optics.\cite{fullager2019metalized, colla2019} Furthermore, polymers compatible with stereolithography were found to be transparent in THz spectral range to potentially enable the fabrication of transmissive optics.\cite{Park2019} Thus, stereolithographic fabrication may open up new pathways for the manufacturing of polymethacrylate-based or integrated THz optical components.}

\red{Polymers are known to undergo morphological changes when heated to temperatures below the melting point, or annealed.\cite{fischer1972effect} Morphological changes are often accompanied by alterations in mechanical and optical properties. Therefore heat-treatment-induced morphological changes have been used to modify optical and mechanical material properties for integrated optics such as fiber Bragg gratings\cite{yuan2011improved, woyessa2016temperature} and polymer-based solar cells, for instance.\cite{erb2005correlation, Li2007solvent} Thus, the thermal stability of polymers is an important aspect to consider when designing an optical component to ensure its performance. However, while thermally induced changes in the mechanical properties of various polymers including PMMA, polycarbonate, and polystyrene, for instance, are well investigated, information on thermal effects in the THz optical properties is still scarce.\cite{brady1971yielding, fischer1972effect, fu2011spectroscopy}}

In this paper, we sought to investigate THz optical response of bulk polymethacrylates subjected to annealing processes. A common polymethacrylate that is compatible with a stereolithographic fabrication has been selected for this investigation. THz spectroscopic ellipsometry was performed before and after the annealing processes. The polymethacrylate studied here was found to maintain its ellipsometric response in the THz spectral range after annealing at moderate temperature of up to 70~$^{\circ}$C. Experimental and best-model calculated THz \red{$\cos(2\Psi)$ and $\sin(2\Psi)\cos(\Delta)$ }spectra are presented. A parametrized model dielectric function composed of oscillators with Gaussian broadening was used for the data analysis and is discussed here.

\section{\label{experiment}Experiment}

\subsection{\label{prep}Sample Preparation}
The samples studied here were prepared using UV-induced polymerization of the ``black'' resin available from Formlabs Inc. For each sample, approximately 2~ml of resin was applied in between two microscope slides positioned parallel to each other on a glass plate. Subsequently, a second glass plate was set on top of the spacers to shape the resin into a thin slab. The assembly was then placed in a UV oven (UVO cleaner model no.~42, Jelight Company Inc.) and was cured for 15 minutes until the resin was fully polymerized. \red{As a result, the finished sample has parallel interfaces with low surface roughness and is suitable for accurate ellipsometric measurements in the THz spectral range. The expected interference oscillations in the THz spectral range due to the plain parallel interfaces will aid in the detection of small changes in the dielectric function, which is a well known interference enhancement approach and frequently used in the visible spectral range.\cite{hilfiker2008survey} } The same procedure was applied to create a total of three identical samples. Once UV polymerized, two samples were thermally annealed in a precision oven at 70~$^{\circ}$C for two different lengths of time, 2~and 4~hours, respectively. For comparison, one sample was not thermally annealed and investigated as a reference.

\subsection{\label{acq}Data Acquisition and Analysis}

The thermally annealed polymethacrylate samples were investigated using a custom-built THz spectroscopic ellipsometer system, which is described in detail in Ref.~\onlinecite{kuehne2018}. The ellipsometer employs a rotating analyzer configuration (polarizer ${-}$ sample ${-}$ rotating analyzer). The instrument is equipped with a backward-wave oscillator source operating in a spectral range from 97 to 179~GHz, which can be extended up to 1010~GHz using GaAs Schottky diode frequency multipliers. A Golay cell was used as a detector. Ellipsometric \red{$\cos(2\Psi)$ and $\sin(2\Psi)\cos(\Delta)$ spectra }were obtained over range from 650 to 950~GHz with a resolution of 5~GHz at two incidence angles: $\Phi _{a}$ = 70$^{\circ}$ and 75$^{\circ}$. 

The un-annealed reference sample was investigated over a wider spectral range using a commercial infrared ellipsometer (Mark I IR-VASE$^{\textregistered}$, J.A. Woollam Company Inc.) and a commercial THz ellipsometer (THz-VASE, J.A. Woollam Company Inc.). The IR ellipsometer operates in a polarizer ${-}$ sample ${-}$ rotating compensator ${-}$ analyzer configuration, while the THz ellipsometer uses a rotating polarizer ${-}$ sample ${-}$ rotating compensator ${-}$ analyzer configuration as detailed in Ref.~\onlinecite{Fujiwara2007}. 

Ellipsometric \red{$\cos(2\Psi)$ and $\sin(2\Psi)\cos(\Delta)$ spectra }were obtained in the infrared spectral range from 300 to 4000~cm$^{-1}$ (9 to 120~THz) with a resolution of 4~cm$^{-1}$ (0.1~THz) at three angles of incidence: $\Phi _{a}$ = 65$^{\circ}$, 70$^{\circ}$, and 75$^{\circ}$. The THz ellipsometric data were obtained over the range from 22 to 32~cm$^{-1}$ (0.65 to 0.95~THz) with a resolution of 0.2~cm$^{-1}$ (5~GHz) at the same angles of incidence as for the infrared data. The optical modeling and data analysis were performed using a commercial ellipsometry data analysis software package (WVASE32$^\textsuperscript{TM}$, J.A.~Woollam Company). \red{The ellipsometric data sets are analyzed using stratified-layer optical model calculations which consist of three layers,  ambient/polymethacrylate/ambient.\cite{JellisonHOE_2004}}   

The ellipsometric data obtained for the reference sample was used to develop a model dielectric function for the polymethacrylate, which is composed of a sum of oscillators with Gaussian broadening: 

\begin{equation}
\label{eq:eps_all}
\varepsilon(\omega) = \varepsilon_{1}(\omega)+i \varepsilon_{2}(\omega) = \varepsilon_{\infty} + \sum_{i} \varepsilon_{Gau}({A}, \Gamma, \omega, \omega_o),
\end{equation}

\noindent where the function $\varepsilon_{\stext{Gau}}({A}, \Gamma, \omega, \omega_o)$ indicates an oscillator with Gaussian broadening. The oscillator amplitude, broadening, and resonance frequency are designated by $A, \Gamma$, $\boldmath \omega_o$, respectively. The oscillators are given analytically by their Gaussian form for the imaginary part $\varepsilon_{2}^{\stext{Gau}}(\omega)$ of the complex dielectric function $\varepsilon(\omega)$:

\begin{equation}
\label{eq:Gaussian}
\varepsilon_{2}^{\stext{Gau}} (\omega) = A e^{\left(-\left(\frac{\omega - \omega_o}{f\cdot\Gamma}\right)^{2}\right)} +  A e^{\left(-\left(\frac{\omega + \omega_o}{f\cdot\Gamma}\right)^{2}\right)},
\end{equation}

\noindent where $1/f = 2\sqrt{\rm{ln}(2)}$. The corresponding values for $\varepsilon_{1}^{\stext{Gau}}(\omega)$ are determined by Kramers-Kronig integration of Eq.~(\ref{eq:Gaussian}) during the Levenberg-Marquardt-based line shape analysis of the experimental spectra. 

\begin{figure}[t]
	\centering
	\includegraphics[width=1\linewidth, trim=0 450 145 0,clip]{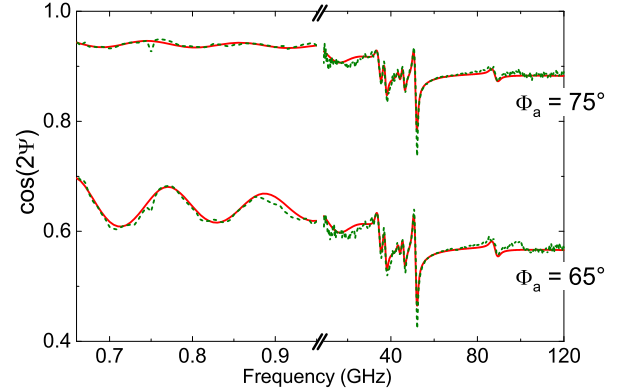}
	\caption{Best-model calculated (solid red lines) and experimental (dashed green lines) \red{$\cos(2\Psi)$ spectra }obtained at $\Phi_{a}$ = 65$^{\circ}$ and 75$^{\circ}$ for the un-annealed reference sample. The infrared range is dominated by a number of distinct absorption bands while the THz range shows Fabry-P\'erot oscillations as a result of the plane parallel interfaces of the sample.}
	\label{fig:psi}      
\end{figure}
\begin{figure}[b]
	\centering
	\includegraphics[width=0.95\linewidth, trim=0 450 160 0,clip]{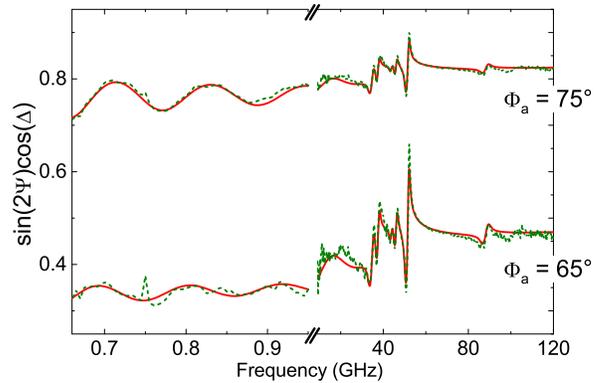}
	\caption{As Fig.~\ref{fig:psi}, but for the best-model calculated (solid red lines) and experimental (dashed green lines) \red{$\sin(2\Psi)\cos(\Delta)$ spectra }obtained at $\Phi_{a}$ = 65$^{\circ}$ and 75$^{\circ}$ for the un-annealed reference sample.}
	\label{fig:delta}      
\end{figure}

\section{\label{result}Results and Discussion}

Figure \ref{fig:psi} and \ref{fig:delta} illustrate the experimental (dashed green lines) and the best-model calculated (solid red lines) $\Psi$- and $\Delta$-spectra of the un-annealed polymethacrylate reference sample, respectively, at two angles of incidence $\Phi_{a}$ = 65$^{\circ}$ and 75$^{\circ}$ for the spectral range from 0.65 to 0.95~THz, and from 9 to 120~THz. 

\begin{figure}[htb]
	\centering
	\includegraphics[width=0.95\linewidth, trim=50 400 60 20,clip]{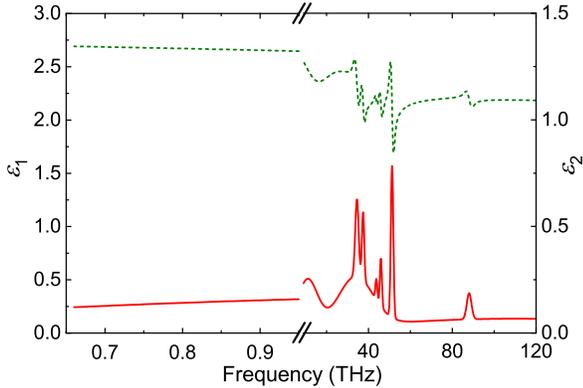}
	\caption{\red{Best-model calculated real (dashed green line) and imaginary part (solid red line) of the complex dielectric function $\varepsilon(\omega)$ for the data shown in Figs.~\ref{fig:psi} and \ref{fig:delta} is depicted. The major absorption features occur in the range from 9 to 120~THz. Below 9~THz only a broad and shallow absorption can be observed. The best-model parameters are omitted here and the interested reader is referred to Ref.~\onlinecite{Park2019}.}}
	\label{fig:dielectric}      
\end{figure}

Note that experimental $\Psi$- and $\Delta$-spectra were obtained for $\Phi _{a} = 65^{\circ}, 70^{\circ},$ and $75^{\circ}$ and analyzed simultaneously, however, for clarity, the data for $\Phi_{a}$ = 70$^{\circ}$ is omitted here.

The experimental and best-model calculated data are in very good agreement for the non-annealed reference sample. Fabry-P\'erot oscillations are observed in the range from 0.65 to 0.95~THz, indicating that the sample is transparent in this spectral window. The IR range (9 to 120~THz), on the other hand, is dominated by a number of distinct absorption features with Gaussian broadening profile. Further details on the best-model parameters are provided in Tabs.~1 and 2 in Ref.~\onlinecite{Park2019}. 

\begin{figure}[tb]
	\centering
	\includegraphics[width=1\linewidth, trim=0 460 150 20,clip]{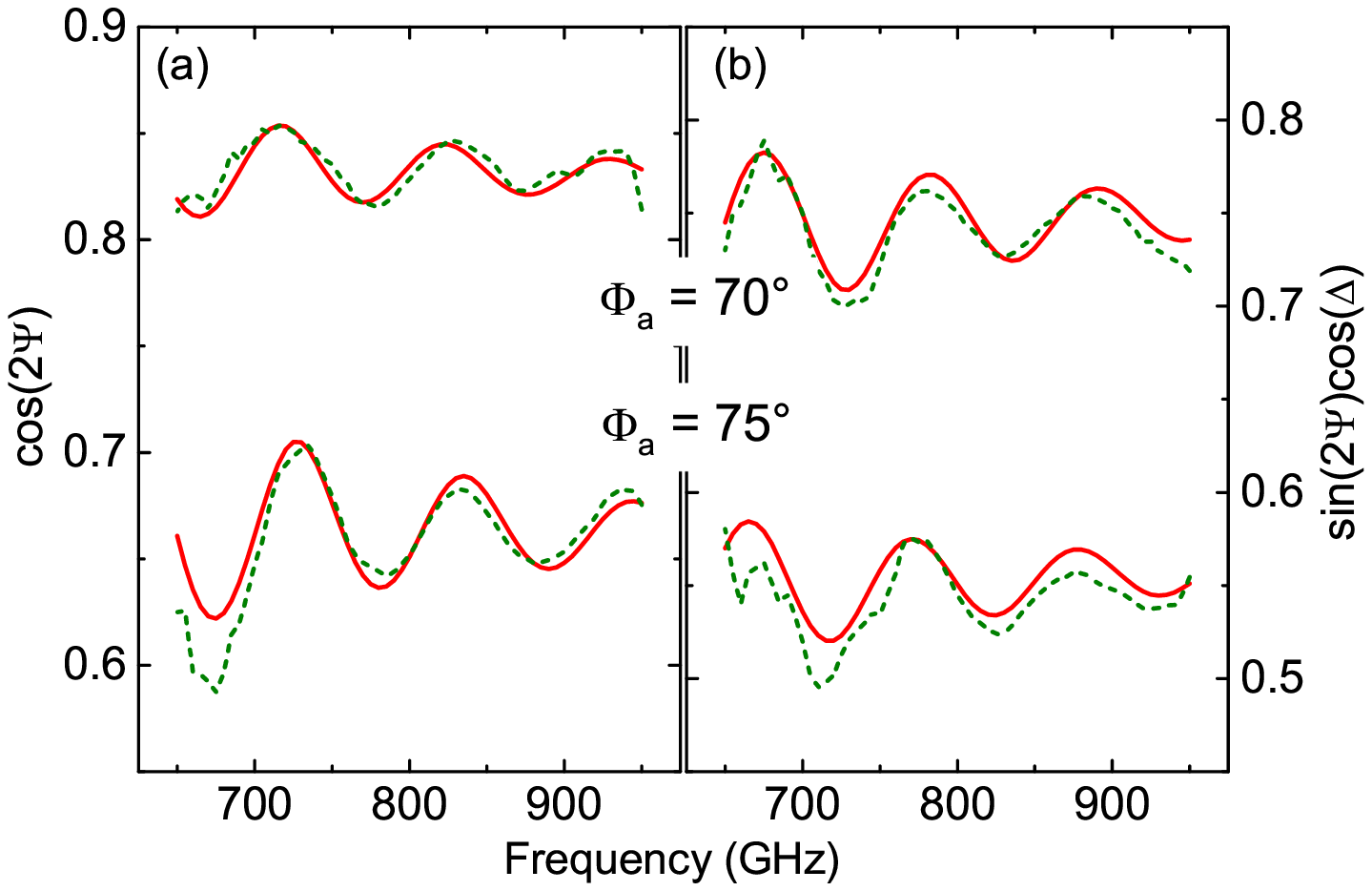}
	\caption{(a) Experimental (green dotted lines) and best-model calculated (red solid lines) \red{$\cos(2\Psi)$ spectra }of the polymethacrylate sample which was annealed for 2~hours obtained at $\Phi _{a}$ = 70$^{\circ}$ and 75$^{\circ}$. Fig.~\ref{2hr}~(b) Same as (a) but for the \red{$\sin(2\Psi)\cos(\Delta)$ spectra.}}
	\label{2hr}      
\end{figure}

\begin{figure}[tb]
	\centering
	\includegraphics[width=1\linewidth, trim=0 460 150 20,clip]{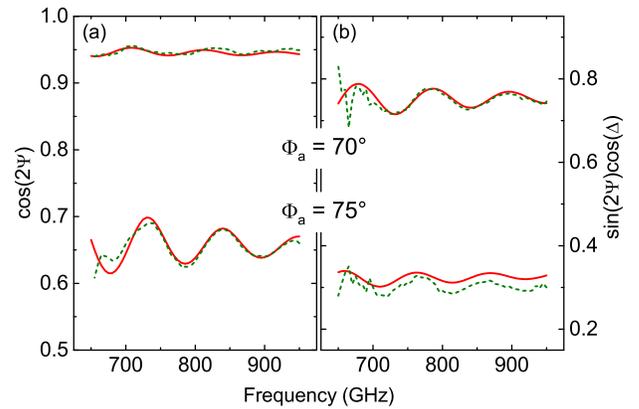}
	\caption{(a) Experimental (green dotted lines) and best-model calculated (red solid lines) \red{$\cos(2\Psi)$ spectra }of the polymethacrylate sample which was annealed for 4~hours obtained at $\Phi _{a}$ = 70$^{\circ}$ and 75$^{\circ}$. Fig.~\ref{4hr}~(b) Same as (a) but for the \red{$\sin(2\Psi)\cos(\Delta)$ spectra.}}
	\label{4hr}      
\end{figure}

Figures \ref{2hr} and \ref{4hr} show the experimental and best-model calculated \red{$\cos(2\Psi)$ and $\sin(2\Psi)\cos(\Delta)$ spectra }obtained for the samples annealed for 2 and 4~hours, respectively. Similar to the ellipsometric data of the reference sample shown in Figs.~\ref{fig:psi} and \ref{fig:delta}, the \red{$\cos(2\Psi)$ and $\sin(2\Psi)\cos(\Delta)$ spectra }of the annealed samples are dominated by Fabry-P\'erot oscillations in spectral range from 650 to 950~GHz. The experimental data obtained for both annealed samples were analyzed using the model dielectric function established for the reference sample without further varying the best-model oscillator parameters. In order to obtain a good agreement between experimental and model-calculated data, only the parameters for the sample thickness were varied. The best-model thicknesses for 2~hours and 4~hours annealed samples were 1080~$\mu$m $\pm$ 1~$\mu$m and 1086~$\mu$m $\pm$ 1~$\mu$m, respectively. These thickness values are in an excellent agreement with the nominal thickness of the samples, which is 1000~$\mu$m. This suggests that annealing for up to 4~hours at a temperature of 70~$^{\circ}$C does not lead to any measurable changes in the THz dielectric function. A slight reduction of the interference amplitude at higher THz frequencies can be noticed in Figs.~\ref{2hr} and \ref{4hr}. We have attributed this reduction to a broad and shallow absorption located outside of the accessible spectral range of the THz ellipsometer at approximately 1.2~THz. For further details regarding this broad absorption, the interested reader is referred to Ref.~\onlinecite{Park2019}. \vspace{0.5cm}

\section{\label{conc}Summary and Conclusion}
A common polymethacrylate compatible with a commercial stereolithography system was investigated before and after thermal annealing by spectroscopic THz ellipsometry. A model dielectric function consisting of Gaussian oscillators was established using THz and infrared ellipsometric data obtained from an un-annealed sample. This model dielectric function was also found to accurately render the experimental \red{$\cos(2\Psi)$ and $\sin(2\Psi)\cos(\Delta)$ spectra }from thermally annealed samples in THz spectral range. In conclusion, the investigated polymethacrylate was found to maintain a stable optical response in THz spectral range after thermal annealing at temperatures up to 70~$^{\circ}$C.

\section{\label{acknowledgement}Acknowledgment}
The authors are grateful for support from the National Science Foundation (1624572) within the I/UCRC Center for Metamaterials, the Swedish Agency for Innovation Systems (2014-04712), and Department of Physics and Optical Science of the University of North Carolina at Charlotte.

\end{document}